\documentclass[prb,nobalancelastpage,twocolumn,superscriptaddress,nofootinbib]{revtex4}
\usepackage{color,amsthm,amsmath,amsfonts,amssymb,graphicx,bm}
\usepackage{ulem}

\def\Id{{\openone}}

\newcommand{\ket}[1]{|#1\rangle}
\newcommand{\bra}[1]{\langle #1|}
\newcommand{\tr}{\text{tr}}
\newcommand{\op}{ \textrm{opt}}
\newcommand{\pf}{\text{pf}}

\newcommand{\opt}{*}

\begin{document}

\title{Robustness of quantum memories based on Majorana zero modes}

\author{L. Mazza}
\email{leonardo.mazza@sns.it}
\affiliation{NEST, Scuola Normale Superiore \& Istituto Nanoscienze-CNR, I-56126 Pisa, Italy}
\affiliation{Max-Planck-Institut f\"{u}r Quantenoptik, Hans-Kopfermann-Strasse 1,
D-85748  Garching, Germany}

\author{M. Rizzi}
\affiliation{Institut f\"{u}r Physik, Johannes-Gutenberg-Universit\"{a}t Mainz, Staudingerweg 7, D-55099  Mainz, Germany}
\affiliation{Max-Planck-Institut f\"{u}r Quantenoptik, Hans-Kopfermann-Strasse 1,
D-85748  Garching, Germany}

\author{M. D. Lukin}
\affiliation{Physics Department, Harvard University, Cambridge, Massachusetts 02138, USA}

\author{J. I. Cirac}
\affiliation{Max-Planck-Institut f\"{u}r Quantenoptik, Hans-Kopfermann-Strasse 1,
D-85748  Garching, Germany}

\begin{abstract}
We analyze the rate at which quantum information encoded in zero-energy Majorana modes is lost in the presence of perturbations. 
We show that information can survive for times that scale exponentially with the size of the chain both in the presence of  quenching and time-dependent quadratic dephasing perturbations, even when the latter have spectral components above the system's energy gap.  
The origin of the robust storage, namely the fact that a sudden quench affects in the same way both parity sectors of the original spectrum, is discussed, together with the memory performance at  finite temperatures and in the presence of particle  exchange with a bath.
\end{abstract}
\maketitle


\section{Introduction}

In the presence of noise and perturbations, the information stored in quantum systems is typically lost. Indeed, orthogonal states (or their superpositions) may evolve into similar states, so that even when the perturbation is known it is impossible to restore the original one.~\cite{NC} 
However, the rate at which information corrupts depends on how it is actually encoded, on the nature of the perturbations, as well as on the properties of the system. 
In particular, in the case of a many-body system, this rate may depend on its size: for the larger Hilbert space there is ``more space'' to keep the states distinguishable. Systems possessing this property for naturally occurring perturbations may serve as quantum memories,~\cite{QMemoExp} which can reliably store quantum states for long times.

In this article, we investigate the capability of a Kitaev  chain~\cite{Kitaev01}  to store quantum states in the presence of noise. 
We consider the encoding of a qubit in the zero-energy Majorana modes of a Kitaev chain described by Hamiltonian $\hat H_0$ and analyze the loss of information occurring when the time evolution describing the storage period is dictated by a different Hamiltonian, $\hat H_0+ \hat V(t)$, where $\hat V(t)$ is an unknown perturbation. 
We explicitly focus on time-dependent quadratic perturbations containing high frequencies and find that the quantum memory can be robust to a wide range of perturbations. Specifically, even though perturbations spread the qubit in the whole Hilbert space and the average over several $\hat V (t)$ is considered, the orthogonality of any pair of initial states is preserved for very long times: the rate at which the information is lost decreases exponentially with the system size.  

The potential use as quantum memories of Kitaev chains~\cite{GoldsteinChamon,QPLosses,DasSarma,BKDisorder} and of other topological systems~\cite{Dennis, Alicki, Terhal, Loss,  Fernando, Fernando2, Alastair, Haah, Hutter} has been extensively studied in recent years, following the seminal work by Kitaev.~\cite{Kitaev,Dennis,Kitaev01}
Most of the prior work has focused on analyzing systems with topological order, whereby the degeneracy of the ground state subspace is stable under small  local perturbations~\cite{Kitaev,Kitaev01,Michalakis} with frequencies well below the characteristic energy gap. This ensures that all of the ground states of the perturbed Hamiltonian have a trivial evolution so that, when encoded in these ground states, the information can survive for long times. 
In contrast, here we consider the situation when highly excited states are created due to sudden  quench perturbations with high-frequency excitations above the gap.~\cite{BKDisorder}
In such cases, conventional topological protection is no longer effective.~\cite{Akhmerov} 

Our work is closely connected to the use of topological systems as error correcting codes in the context of self-protected quantum memories.~\cite{NC, Dennis, Alicki, Terhal, Loss,  Fernando, Fernando2, Alastair, BKDisorder, Haah, Hutter} However, in contrast to previous investigations, we do not consider a specific error correction procedure,~\cite{Gottesmann} but rather analyze whether the information is still present ``somewhere'' in the Hilbert space. In fact, our analysis shows that the memory time
grows exponentially with the system size in situations where previous approaches give a negative result. Besides, our results may bear an interesting connection to the phenomenon of many-body localization~\cite{Altshuler, Huse}, whereby different initial states remain distinguishable for arbitrarily long times under a quench.

In this article we introduce a series of theoretical tools
to analyze the loss of  quantum information encoded in many-body systems and 
characterize the optimal procedure to decode it. 
In a quantum-memory framework, we quantify the amount of information that may be recovered after the storage period and evaluate it, 
as well as its upper and lower bounds, for different relevant cases. 
For the problem of interest, our numerical methods can investigate relatively large systems, and the appropriate size-scalings are derived. 
We identify the conditions for a memory time that scales exponentially with the system size and study how this exponential dependence is affected by finite  temperatures, or  by the particle  exchange between the chain and a reservoir. 

This article is organized as follows.
In Sec.~\ref{sec:setup} we review the theory of a quantum memory based on the Majorana modes of two Kitaev chains, and set up the notation.
In Sec.~\ref{sec:rec:of:info} we present the theory of the optimal recoverability of the information, both for general and for Gaussian recovery operations.
In Sec.~\ref{sec:result} we present the main result of the article, namely the fact that a quantum memory based on the Kitaev chain can withstand a sudden quench perturbation.
In Sec.~\ref{sec:discussion} we further elaborate on it, and the effect of temperature and particle losses is considered.
Finally, in Sec.~\ref{sec:conc} we present our conclusions.

\section{Setup}\label{sec:setup}

We consider a Kitaev chain of $2N$ Majorana (real) modes, $\{ \hat c_j \}$, with $\hat c_j^\dagger = \hat c_j$ and $\{\hat c_j, \hat c_k \}_+ = 2 \delta_{j,k}$.
The Hamiltonian is:~\cite{Kitaev01}
\begin{align}
 \hat H(\mu,\Delta) =& \frac{ i J }2 \sum_{j=1}^{N-1} \hat c_{2j} \hat c_{2j+1} - \frac {i\mu}2 \sum_{j=1}^{N} \hat c_{2j-1} \hat c_{2j}
 + \nonumber \\
 &+i\frac{|\Delta|-J}{2}\sum_{j=1}^{N-1}\hat c_{2j-1}\hat c_{2j+2}.
 \label{eq:protecting}
\end{align}
At zero temperature, it has a topological phase for $|\mu/J| < 2, \Delta \neq 0$ and the ground state is quasi-degenerate due to the existence of (nearly) zero energy Majorana modes, $\hat m_{1,2}$, localized at the edges. 
They can be expressed as linear combinations of the $\{\hat c_j\}$, and can be used to define a Dirac mode $\hat a = \frac 12 \left( \hat m_{1} + i \hat m_{2} \right)$. 
Due to superselection rules on the parity of the number of fermions, 
two such Dirac modes are necessary in order to define a meaningful qubit.
Thus, we consider a second Kitaev chain, whose zero-energy Majorana modes are $\hat m_{3,4}$, and which define a Dirac mode $\hat b = \frac 12 (\hat m_3 + i \hat m_4)$. 
We define a qubit in the even parity sector of the model: $\text{span}\{ \ket{0}=\ket{\mathrm{vac}}, \ket{1} =\hat a^\dagger \hat b^\dagger \ket{\mathrm{ vac}} \}$
and construct a set of Pauli operators:
\begin{equation}
\hat{\sigma}'_x = -(\hat a \hat b + \hat b^{\dagger} \hat a^{\dagger}); \;\;
\hat \sigma'_y = i(\hat a \hat b - \hat b^{\dagger} \hat a^{\dagger}); \;\;
\hat \sigma'_z = \hat 1 - \hat a^{\dagger} \hat a - \hat b^{\dagger} \hat b.
\label{eq:paulitilde}
\end{equation}
Results would be identical, apart from notation, in the odd sector: $\text{span}\{ \ket{\tilde 0}= \hat a ^\dagger \ket{\mathrm{ vac}},
\ket{\tilde 1} = \hat b^\dagger \ket{\mathrm{vac}} \}$.

\begin{figure}[t]
\includegraphics[width=\columnwidth]{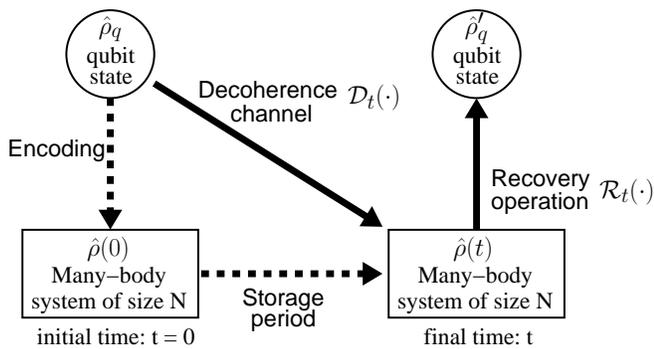}
\caption{A qubit $\hat \rho_q$ is encoded into a topological system of size $N$ and stored for a time $t$. Perturbations act on the system during the storage and induce a non-unitary time-evolution that
destroys the encoded information. We model this process with a decoherence channel $\mathcal D_t(\cdot)$. A recovery operation attempts at retrieving the information and reconstructs a qubit $\hat \rho_q'$ which should be as similar as possible to $\hat \rho_q$.}
\label{fig:sketch}
\end{figure}

When the Hamiltonian is perturbed or there is an interaction with the environment, the chain undergoes a non-trivial dynamics. Let us denote by ${\cal D}_t$ the {\it decoherence channel} that describes the encoding of a qubit into the ground state of the chains and the subsequent time-evolution which describes the storage of the information (see Fig.~\ref{fig:sketch}).
Note that the latter takes into account both the Hamiltonian part of the time evolution and the non-unitary action of perturbations.

The decoherence channel maps the state of the qubit onto the state of the two chains;
therefore, ${\cal D}_t( \hat\rho_q )= \hat\rho(t)$, where $\hat\rho_q$ is a qubit density operator and $\hat\rho(t)$ is the state of the chains at time $t$.
$\hat \rho (0)$ thus represents the many-body state where the information is encoded and it must always fulfill $\langle\hat \sigma_\alpha \rangle_{\hat\rho_q} =\langle \hat\sigma_\alpha'\rangle_{\hat\rho(0)}$, where $\{ \hat \sigma_\alpha \}$ are the usual Pauli matrices and the $\{ \hat\sigma_\alpha' \}$ are defined in~\eqref{eq:paulitilde}.
These conditions do not define $\hat \rho (0)$ uniquely; where not explicitly stated, we will consider:
\begin{equation}
 \hat \rho (0) = \frac 1 {\mathcal N} 
 \left( \hat{\mathbb I} + 
 \sum_\alpha \langle\hat \sigma_\alpha \rangle_{\hat\rho_q} \hat \sigma_\alpha'
 \right) 
 \left(
 \prod_{\gamma>0} \hat f_\gamma \hat f_\gamma^{\dagger}
 \right)
\label{eq:encoding:1}
\end{equation}
where operators $\hat f_{\gamma}^{(\dagger)}$ diagonalize the total Hamiltonian and $\gamma>0$ labels those modes which have non-zero energy; $\mathcal N$ is a properly chosen constant.

In order to simplify the discussion, we assume that the chains are kept far apart,  so that they  do not interact  with each other or with the same environment. Moreover, throughout the article we assume that the first Dirac  mode, $\hat a$, does not undergo any dynamics, neither Hamiltonian nor dissipative, as in general  only the relative time evolution of the two chains contributes to the corruption of the information.
In this setting there is no need to keep track of the modes of the first chain different from $\hat a$.

\section{Optimal recovery of the information}\label{sec:rec:of:info}

After the storage time $t$, the attempt to retrieve the initial qubit is described by  the \textit{recovery channel}, ${\cal R}_t$, which maps back the chain and the extra fermionic mode, $\hat a$, to a qubit 
(see Fig.~\ref{fig:sketch}).
It has to be chosen such that the composite channel ${\cal T}_t \doteqdot {\cal R}_t\circ {\cal D}_t$ is as close as possible to the identity channel. This can be quantified with the \textit{recovery fidelity}:~\cite{Bowdrey}
\begin{equation}
 F({\cal R}_t) = \int d\mu_\varphi \langle\varphi| \, {\cal T}_t
 \left(|\varphi\rangle\langle\varphi| \right)  |\varphi\rangle,
 \label{eq:F:average}
\end{equation}
where the integral is over the surface of the Bloch sphere.

In Appendix~\ref{appendix1} we demonstrate that, given a decoherence channel, the optimal fidelity $F^{\op}_t$ is given by
 \begin{equation}
 \label{eq:Fopt}
 F^{\op}_t = \max_{\mathcal R_t} F(\mathcal R_t)=\frac{2}{3}+\frac{1}{6} \| \hat \rho_{+}(t)- \hat \rho_{-}(t) \|_{\tr};
 \end{equation}
where $\hat \rho_{\pm }(t) \equiv \hat \rho_{x, \pm }(t) = \mathcal D_t (\hat \Psi_{x,\pm})$,
$\hat \Psi_{x,\pm } = ( \hat 1 \pm \hat \sigma_x)/2$
are the (pure) eigenstates of $\hat \sigma_x$ with eigenvalues $\pm1$, 
and $\|~\cdot~\|_\tr$ denotes the trace norm, i.e. the sum of the singular values of the operator.
Eq.~\eqref{eq:Fopt} shows that $F^{\op}_t$ depends on the distance 
$d(\hat \rho, \hat \omega) \doteqdot  \| \hat \rho - \hat \omega \|_{\text tr} / 2$ 
between two many-body states which result from the time evolution of two orthogonal qubit states.
At $t=0$, $d (\hat \rho_{+ }(0),\hat \rho_{- }(0)) = 1$ but it 
decreases with time
due to noise and perturbations, and no physical operation can increase it 
again because of the contractivity of the trace norm.~\cite{NC}

In Appendix~\ref{appendix1} we also characterize the properties of the \textit{optimal recovery operation} ${\cal R}^{\op}_t$
that saturates the bound $F_t^{\op}$ of Eq.~\eqref{eq:Fopt}. 
This ${\cal R}^{\op}_t$ consists in the measurement on $\hat \rho(t)$ of three many-body observables $\hat H_\alpha$,
that (i) essentially undo the rotation that the axes of the original qubit $\hat{\rho}_q$ have undergone,
and (ii) quantify the distance between pairs of initially orthogonal states aligned on them.
Indeed, they are chosen so that $\tr[ \hat H_\alpha \mathcal D_t (\hat \sigma_\alpha)]  \doteqdot \| \mathcal D_t (\hat \sigma_\alpha)\|_\tr
= 2 d( \hat \rho_{\alpha,+}(t) , \hat \rho_{\alpha,-}(t) )$,
where we applied the linearity of $\mathcal D_t$ on the eigenstates
$\hat \Psi_{\alpha,\pm} = (\hat I \pm \hat \sigma_\alpha)/ 2$ of $\hat \sigma_\alpha$:
\begin{equation}
\label{eq:Recov}
{\cal R}^{\op}_t(\hat \rho(t)) = 
\frac{1}{2} \hat {\mathbb I} \, \tr\left[\hat \rho (t) \right] + \frac{1}{2} \sum_\alpha \hat \sigma_\alpha \,\tr\left[\hat H_\alpha\hat \rho (t) \right].
\end{equation}
Alternatively, the optimal recovery operation can be thought of as a two-step procedure:
the first being a unitary acting over the global Hilbert space that attempts
to align back the image of the Bloch sphere,
and the second a trace on all 
the fermionic modes different from $\hat a$ and $\hat b$, where the initial qubit was defined and the recovered one is recreated.
Notice that the unitary (defined in terms of $\hat H_\alpha$'s) may involve $N$-body terms acting on all the fermionic modes of the chain
and the specific form depends on the singular value decomposition of $\hat \rho_+(t)-\hat \rho_-(t)$.

Because the optimal recovery operation can be difficult to implement in practice, it is natural to restrict the optimization of $F({\cal R}_t(\cdot))$ to those physical actions that can be operatively realized.
Those actions depend on the specific experimental setup and should be independently studied for each physical implementation. 
Here we consider as experimentally-relevant recovery operations those which are fermionic Gaussian channels.~\cite{FGS, FGDissipative} 
They comprise the addition and discard of auxiliary modes, the time evolution under quadratic Hamiltonians and under master equations with linear jump operators (see Appendix~\ref{sec:FGstates} for a short introduction).
In Appendix~\ref{appendix2} we show that
the best recovery fidelity attainable with a fermionic Gaussian channel is:
\begin{equation}
F^{\op}_{\mathrm G, t} = 
\max_{\mathcal R_t(\cdot)  \text{ is Gaussian}} \hspace{-0.15cm}F(\mathcal R_t)
=
\frac{2}{3}+
\frac{1}{6} \|\Gamma_{+}(t)- \Gamma_{-}(t)\|_{\rm op}.
\label{eq:FGopt}
\end{equation}
Here, $\Gamma_{\pm,}(t)$ are the covariance matrices (CM) of $\hat \rho_\pm (t)$, defined as $[\Gamma_{\pm}(t) ]_{j,k} \doteqdot i \text{tr} [\hat \rho_\pm (t) \hat c_j \hat c_k]$.
The $\| \cdot \|_{\rm op}$ denotes the operator norm, i.e. the largest singular value of the operator.
Similarly to the previous case, it is possible to characterize the properties of the \textit{optimal Gaussian recovery operation} that achieves $F^{\op}_{\mathrm G, t} $  (see Appendix~\ref{appendix2}).

\section{Main Results}\label{sec:result}

\begin{figure}[t]
\begin{center}
\includegraphics[width=\columnwidth]{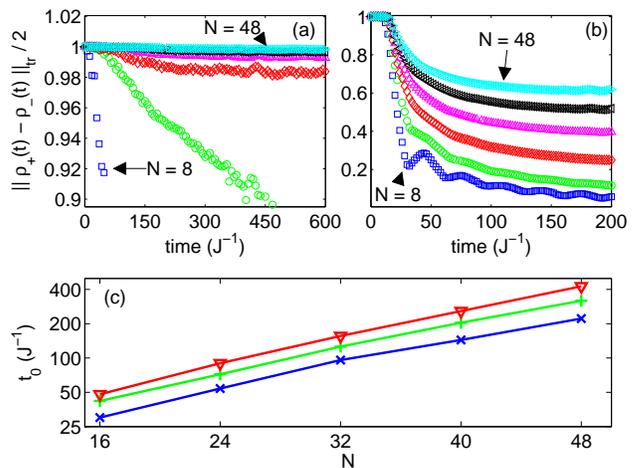}
\end{center}
\caption{Robustness of the quantum memory against Hamiltonian perturbations:  $\mu_0=0$, $\Delta_0 = J$ and: (a,c) $\mu_-=J$, $\mu_+=1.5 J$; (b) $\mu_-=2.5J$, $\mu_+=3 J$. Upper panels (a,b): Fidelity as a function of time for different chain lengths, $N= 8$, $16$, $24$, $32$, $40$ and $48$. Lower panel (c): Memory time $t_0$ as function of $N$ for different values of the  fidelity threshold $F_0$: 
$F_0 = 0.9995$ blue line (x marker), $F_0 = 0.999$ green line (+ marker), $F_0 = 0.9985$ red line (triangular marker).
We considered $N_{\rm d}= 101$ realizations uniformly distributed in the range $[\mu_-,\mu_+]$.
Data show convergence behavior for $N_{\rm d} \to \infty$ (see Appendix~\ref{app:Ham:pert}) and can therefore be considered as an approximation of the continuum situation.
}
\label{fig:decomap:unknownHam}
\end{figure}

We employ Eq.~\eqref{eq:Fopt} to show that the qubit can efficiently withstand sudden changes of the Hamiltonian during the storage period.
First, we assume the perfect encoding of the qubit using the two lowest-energy eigenstates of $\hat H_0 \doteqdot \hat H(\mu_0,\Delta_0)$. 
For the storage, a perturbation, $\hat V$, randomly chosen from a set according to a measure $\nu_{\hat V}$, is added to $\hat H_0$: the global Hamiltonian describes the time evolution.
Finally, the resulting states are incoherently added ($\hbar = 1$):
\begin{equation}
\mathcal D_t(\hat \rho_q) = \int \mathrm d \nu_{\hat V} \;
e^{-i (\hat H_{0} + \hat V) t} \hat \rho(0)
e^{i (\hat H_{0} + \hat V) t}.
\label{eq:decomap:unknownHam}
\end{equation}
We choose $\hat V$ quadratic in the Majorana operators, so that each term of the integral can be efficiently computed. Moreover,  we consider
a discrete measure and exploit the fact that $\mathcal D_t(\hat \rho_q)$ has support in a relatively small subspace (see Appendix~\ref{app:Ham:pert} for more details on the calculation).

\begin{figure}[t]
\begin{center}
\includegraphics[width=\columnwidth]{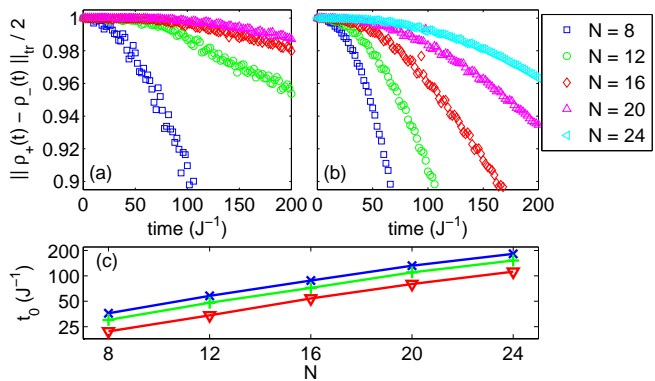}
\end{center}
\caption{
Robustness of the quantum memory against time-dependent Hamiltonian perturbations. Panel (a): $\mu_0=0$, $\Delta_0 = J$ and $\mu_-=J$, $\mu_+=1.5J$. The Hamiltonian time-evolution is swapped between $\hat H(\mu_-)$ and $\hat H(\mu_+)$ every $ \delta t= J^{-1}/4$. $N_{\rm d} = 101$.
Panel (b):  $\mu_0=0$, $\Delta_0 = 2 J$.
The parameters of $\hat H_0 + \hat V$ oscillate as: $\mu = \bar \mu + \delta \mu \cdot \text{sgn}[\sin (2 \pi \omega t)]$ and $\Delta = \bar \Delta + \delta \Delta \cdot \text{sgn} [\sin (2 \pi \omega t)]$, $\omega = 2J$, $\bar \mu = 1.1 J$, $\bar \Delta = 1.9 J$. We consider $N_{\rm d} = 144$ realizations, with $\delta \mu$ and $\delta \Delta$ uniformly distributed in the ``square'' regions $\delta \mu \in [-0.1 J, 0.1 J]$ and $\delta \Delta \in [-0.1 J, 0.1 J]$. 
For every realization we add a (different) site-dependent and time-dependent randomness in the chemical potential $\mu_i (t) = \mu'_i \left(\frac{1}{2}+\frac{1}{2} \text{sgn}[\sin(2 \pi \omega t)]\right) +  \mu''_i \left(\frac{1}{2}+\frac{1}{2} \text{sgn}[-\sin(2 \pi \omega t)]\right)$. $\mu_i'$ and $\mu_i''$ are taken randomly with a flat distribution in $[-0.1 J, 0.1J]$.
Panels show the optimal fidelity as a function of time for different sizes of the chain, $N$. 
Panel (c): Memory time $t_0$ as function of $N$ for different values of the fidelity threshold $F_0$ for the case of panel (b): $F_0 = 0.97$ blue line (x marker), $F_0 = 0.98$ green line (+ marker), $F_0 = 0.99$ red line (triangular marker).
}
\label{fig:decomap:unknownHam2}
\end{figure}

We begin considering a quenching perturbation in the chemical potential of the system. For times $t>0$ we consider the evolution with  $\hat H_0+ \hat V \doteqdot \hat H (\mu\neq\mu_0,\Delta_0)$, representing a classical uncertainty in the number of particles, or the coupling to an unshielded field. We take $N_{\rm d}$ values for $\mu$ uniformly distributed in $[\mu_-,\mu_+]$. Figure \ref{fig:decomap:unknownHam}(a,b) shows $F^\op_t$ as a function of $t$ for different system sizes. $\mu_\pm$ is chosen so that the perturbed Hamiltonians lie inside (a) and outside (b) the topological phase, respectively. In the former case the decay time of $F^\op_t$ strongly depends on the size $N$, whereas in the latter case the size-dependence is  weaker, especially at short times.
Panel (c) displays the memory time,~\cite{BKDisorder} $t_0(N)$, the time at which a prescribed fidelity threshold, $F_0$, is crossed, versus $N$ for the data in (a).
$t_0(N)$ is computed intercepting $F_0$ with a polynomial interpolation of $F_t^\op$ in order to ignore the effect of fast unessential oscillations. Results are compatible with an exponential growth of the memory time with $N$.

Time-independent perturbations have been recently considered also in Ref.~[\onlinecite{BKDisorder}], where it is shown that the memory time does not grow exponentially with the system size.
The discrepancy with our results originates from the choice of the syndrome based recovery operation,~\cite{Gottesmann} which, according to our analysis, is not the optimal one. Nevertheless, when a random site-dependent chemical potential is added, excitations are localized and the memory time increases (see also Ref.~[\onlinecite{Pachos, Pollett}]). Our analysis shows that, even if the standard techniques to correct errors fail, the qubit is still intact and can be recovered with the optimal recovery operation. 

The stability of the memory can be traced back  to the fact that for any pair of Hamiltonians $\hat H_{a,b} \doteqdot \hat H_0+ \hat V_{a,b}$ it holds that:
 \begin{equation}
 \label{Eq:Condition}
 \bra{0} e^{i \hat H_a t} e^{-i \hat H_b t} \ket{0} \approx 
 \bra{1} e^{i \hat H_a t} e^{-i \hat H_b t} \ket{1},
 \end{equation}
where $\ket 0$ and $\ket 1$
are the two ground states for the original Hamiltonian, $H_0$,
and $\approx$ denotes equality apart for a term decaying exponentially with $N$.
Indeed, in Appendix~\ref{app:Ham:pert} we define 
the $N_{\rm d} \times N_{\rm d}$  matrices $G_0$ and $G_1$:
\begin{equation}
 [G_\tau]_{j,k} = 
 \bra{\tau} e^{i  \hat H_j t}
e^{-i \hat H_kt} \ket{\tau}, \; \tau \in \{0,1\};
\quad \hat H_j \doteqdot \hat H_{0}  + \hat V_j.
\end{equation}
and show that:
\begin{equation}
\label{eq:thesis1}
\frac{1}{2} \left\| \hat \rho_{+}(t) - \hat \rho_{-}(t) \right\|_\tr = 
\left \langle \sqrt{G_0/N_{\rm d}} , \sqrt{G_1/N_{\rm d}}
\right \rangle
\end{equation}
where $\langle \cdot , \cdot \rangle$ is the Hilbert-Schmidt inner product for matrices: $\langle A,B \rangle = \tr \left[ B^\dagger A\right]$.
It follows that perfect recoverability, namely $ \left\| \hat \rho_{x,+}(t) - \hat \rho_{x,-}(t) \right\|_\tr=2$, is equivalent to $G_0 = G_1 $.
This means that the excitations generated by any pair of random perturbations have almost the same overlap (in modulus and phase) independent on whether they originated from the $\ket 0$ or $\ket 1$ states. 
The relation between Eq.~\eqref{Eq:Condition} and the topological nature of the states  
$\ket 0$ and $\ket 1$ and of the perturbed Hamiltonians is to be further investigated, as well as its extrapolation to other scenarios.
For a further discussion see Appendix~\ref{app:Ham:pert}.

We next show that these results extend to the case in which $\hat V (t)$ is time-dependent and spatially inhomogeneous. Fig.~\ref{fig:decomap:unknownHam2} shows
$F_t^\op$ as a function of time when the parameters of $\hat V (t)$ oscillate according to a square wave: $\text{sgn}[\sin(2 \pi \omega t)]$ and $\omega = 2J$. 
In panel (a) a homogeneous (global) perturbation of the chemical potential is considered and the memory time appears to increase with the system size. 
In order to show that this behavior is not resulting from a fine-tuned choice of the parameters,
in panel (b) we include both global and site-dependent perturbations in the chemical potential and the pairing term, as well as we consider a different initial state ($\mu_0 =0$, $\Delta_0=2J$). Panel (c) shows that these results are compatible with an exponential growth of the memory time with $N$, whose time-scale depends on the specific situation considered.
Notice that we have chosen an oscillation frequency well above the Hamiltonian gap of $\hat H_0$, of the order of $J$. Thus we conclude that the observed behavior is not directly related to the ground state properties of  Hamiltonian $\hat H_0$. 

\section{Discussion} \label{sec:discussion}

\begin{figure}[t]
\includegraphics[width=\columnwidth]{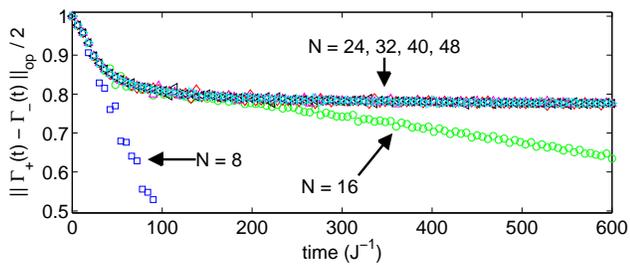}
\caption{Optimal Gaussian fidelity as a function of time for different system sizes $N$: $\mu_0=0$ and $\mu_-=J$, $\mu_+=1.5 J$.
We considered at most  $N_{\rm d}= 101$ realizations uniformly distributed in the range $[\mu_-,\mu_+]$; plotted values for $N_{\rm d} \to \infty $ are obtained via scaling as $1/N_{\rm d}$ (see Appendix~\ref{app:Ham:pert}).
}
\label{fig:GaussianRecovery}
\end{figure}

\subsection{Prior Knowledge of the Recovery Operation}\label{sec:prior}

From the previous discussion, it may appear that the exact distribution of the parameters corresponding
to the possible perturbations is necessary in order to implement the ``correct'' best recovery operation,
as formulated in Eq.~\eqref{eq:Recov}.
Interestingly, instead, we can show that only a coarse knowledge of the interval $I_1$ of possible values is sufficient,
as long as it includes the real one $I_2 \subset I_1$, to obtain a recovery whose quality (in terms of fidelity or storage time)
increments exponentially with the size $N$ of the physical system. 

For the sake of simplicity, we consider the case where only one parameter (e.g., the chemical potential $\mu$)
varies according to box distributions in $I_1$ and $I_2$,
but generalizations on both hypothesis would be straightforward.
Let us call $\mathcal D^{(1)}_t$ and $\mathcal D^{(2)}_t$ the decoherence channels associated with the noise model 
as in Eq.~\eqref{eq:decomap:unknownHam}, 
$\hat \rho_1 \doteqdot \mathcal D^{(1)}_t (\hat \rho_q)$ and $\hat \rho_2 \doteqdot \mathcal D^{(2)}_t (\hat \rho_q)$
their outcomes descending from the pure state input $\hat \rho_q$,
and $\mathcal R_t^{\op,(1)}$ the absolute best recovery operation for $\mathcal{D}_t^{(1)}$
computed by making use of Eq.~\eqref{eq:Recov}.
By definition, we know that  $\| \mathcal R_t^{\op,(1)} (\hat \rho_1) - \hat \rho_q \| < \varepsilon$, 
with $\varepsilon$ scaling exponentially with the size $N$.
In Appendix~\ref{app:prior} we show that $ \|  \mathcal R_t^{\op,(1)} (\hat \rho_2) - \hat \rho_q \| \leq  2\sqrt{\varepsilon/p}$, 
 where $p= |I_2|/|I_1|$, a bound that also scales exponentially with the size $N$.
Thus, the application of $\mathcal R_t^{\op,(1)}$ after the action of $\mathcal D^{(2)}$ 
defines a memory time that, though not absolute best, still increases exponentially with the size of the system.
It is therefore demonstrated that one does not need a fine-tuned knowledge of the perturbation
(and therefore of the absolute optimal $\mathcal R_t^{\op,(2)}$) 
in order to gain exponential protection of the qubit storage.

\subsection{Interactions}

So far, we considered only perturbations preserving the quadratic character of the Hamiltonian. It would be interesting to analyze whether the exponential scaling persists even in the presence of more general perturbations, and in particular of interactions. 
Although such a study is outside the scope of the present work  because it requires the use of different numerical techniques, we believe that interactions may not lead to a dramatic change of the picture arising from the previous discussion. 
Indeed, time evolution in presence of interactions can be considered within the generalized Bogoliubov theory.~\cite{Kraus}  
The result is a non-linear problem that can be mapped onto a quadratic time-dependent Hamiltonian 
whose coefficients depend on the state which is evolved in time.
Remarkably, such time-dependent Hamiltonian 
has the form of those investigated above, 
so that we still expect the memory time to increase with the system size.

\begin{figure}[t]
\includegraphics[width=\columnwidth]{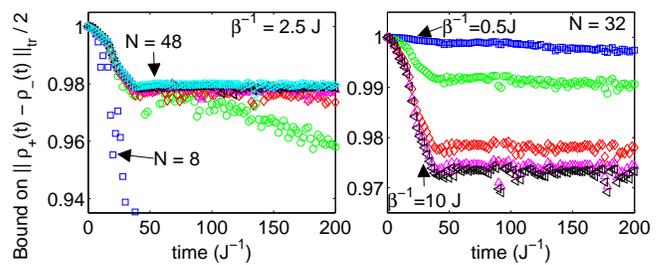}
\caption{Upper bound to the optimal fidelity in presence of thermal modes. (left) Temperature is fixed: $\beta^{-1}=2.5 J$ and the figure shows the upper bound as a function of time for different lengths of the chain $N = 8$, $16$, $24$, $32$, $40$, $48$.
(right) Size is fixed: $N = 32$ and the figure shows the upper bound as a function of time for different temperatures  $\beta^{-1} = 0.5 J$, $1.0 J$, $2.5 J$, $5 J$ and $10 J$.
We take $N_{\rm d}=101$ realizations.
}
\label{fig:decomap:temperature}
\end{figure}

\subsection{Gaussian Recovery Operations and Temperature}

We next examine memory performance for the restricted set of  Gaussian recovery operations. Fig.~\ref{fig:GaussianRecovery} displays the optimal Gaussian fidelity in the presence of the same perturbations as in Fig.~\ref{fig:decomap:unknownHam}(a): the fidelity improves with the size of the system but saturates for sufficiently large $N$. 
Remarkably, similar results are obtained when the encoding is performed in the state of the second chain where all of the non-zero modes are initially in thermal equilibrium at a temperature $\beta^{-1}$:
\begin{equation}
 \hat \rho_\beta (0) = \frac 1 {\mathcal N'} 
 \left( \hat{\mathbb I} + 
 \sum_\alpha \langle\hat \sigma_\alpha \rangle_{\hat\rho_q} \hat \sigma_\alpha'
 \right) 
 \frac{e^{- \beta \sum_{\gamma > 0} \varepsilon_\gamma \hat f_\gamma^\dagger \hat f_\gamma}}{\text{tr}[e^{- \beta \sum_{\gamma > 0} \varepsilon_\gamma \hat f^\dagger_\gamma \hat f_\gamma}]}.
\end{equation}
Operators $\hat f_{\gamma}$ diagonalize $\hat H_0$ and $\varepsilon_\gamma$ is the corresponding energy; $\varepsilon_0 \sim e^{-N}$ and $\hat f_0 \equiv \hat b$.

Because we deal with convex combinations of mixed states, 
we can only plot an upper bound $F^{\rm up}_t$ to the optimal fidelity~(see appendix~\ref{app:Ham:pert}), which we observed to be very close to the exact value in cases in which the latter was computable (not shown).
For temperatures above the gap, Fig.~\ref{fig:decomap:temperature}~(left) displays a clear saturation behavior. For temperatures below the gap, our results are not conclusive (not shown), even if lowering the temperature keeping a fixed value of $N$ clearly increases the saturation bound, as shown in Fig.~\ref{fig:decomap:temperature}~(right). 
Thus, it appears that the temperature  defines an effective system size up to which topological protection can occur.
These results can be understood as follows: (i) although each term in the integral~\eqref{eq:decomap:unknownHam} is Gaussian, its sum (convex combination) is not, and thus the density operator is not Gaussian either; (ii) a Gaussian recovery operation can only depend on the CM of the states $\mathcal D_t (\hat \rho_q)$, which for (pure) non-Gaussian states coincides with that of a mixed Gaussian state.~\cite{footnote3} Thus, restricting to Gaussian recoveries has the same effect as considering mixed states, and this explains the similarity between this case and that of finite temperature.

\subsection{Markovian Interaction with a Bath}

Finally, we consider the effect of particle interchange  with the environment. Previous works already indicate that a quantum
memory may be extremely sensitive to any such perturbation.~\cite{QPLosses} We find that this remains the case even when the optimal recovery is used. 
To show this,  we describe the interaction of the system with a bath with which it can interchange particles via a Lindblad master equation:
\begin{equation}
\partial_t \hat{\rho}  = - i \left[ \hat H_0 , \hat{\rho}  \right] +
\Gamma \sum_{n=1}^N \left(
\hat d_{n} \, \hat{\rho} \, \hat d_{n}^{\dagger} - \frac 12 \{ \hat d_{n}^{\dagger} \hat d_n  , \hat{\rho} \}_+ \right).
\label{eq:master:equation}
\end{equation}
Here $\hat d_n=\frac 12 (\hat c_{2n-1}+i \hat c_{2n})$ annihilates one fermion in the $n$-th physical site, and $\hat H_0$ is the protecting Hamiltonian. This equation transforms Gaussian states into Gaussian states, and thus can be rewritten in terms of the CM.~\cite{FGS,FGDissipative} The optimal fidelity~\eqref{eq:Fopt} cannot be directly computed in terms of the CM. In order to circumvent this problem, we bound $F^\op_t$ with $\frac 12 \| \hat \rho_{x,+}(t) - \hat \rho_{x,-}(t)\|_\tr \leq ( 1 - F_U(\hat \rho_{x,+}, \hat \rho_{x,-})^2)^{1/2}$, where $F_U$ is the Uhlmann fidelity,~\cite{NC} which for Gaussian states is a function of their CM 
(see Appendix~\ref{sec:FGstates}). 
Results in Fig.~\ref{fig:poisoning} demonstrate that the information is corrupted at a rate which does not depend on the system size, the reason being the uniqueness of the fixed point of~\eqref{eq:master:equation}, reached in a time independent of the size.

\begin{figure}[t]
\includegraphics[width=\columnwidth]{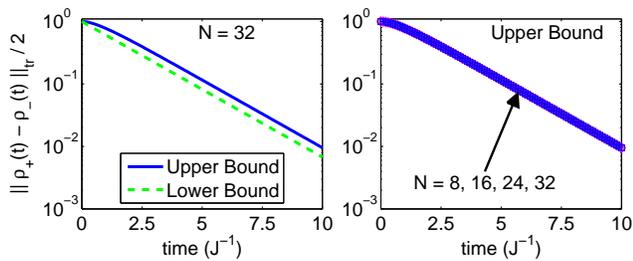}
\caption{Effects of particle losses according to equation~\eqref{eq:master:equation}; $\mu_0=0$ and $\Gamma=J$. (left) Upper and lower bound to the optimal fidelity as a function of time, showing a clear exponential scaling. (right) Upper bound to the optimal fidelity as a function of time for different system sizes: $N=8$, $16$, $24$ and $32$. Results coincide exactly and are indistinguishable.}
\label{fig:poisoning}
\end{figure}

\section{Conclusions}\label{sec:conc}

Our work demonstrates that quantum information stored in Kitaev's chain
can be robust to  perturbations if the optimal recovery of information is used. 
We show that for a broad range of both quenching and time-dependent perturbations the memory time scales exponentially with the system size, and we have identified the condition~\eqref{Eq:Condition} which is responsible for this result. This effect disappears at sufficiently large temperatures, as well as when the system interchanges particles with a bath. 

Our results open a number of interesting research directions. 
First, it would be interesting to explore in detail the stability of the memory when interactions are included. This question could be answered for relatively small chains with an exact numerical calculations. Because the study of large systems and long times may be beyond current numerical possibilities, an experimental quantum simulation might be required.  
Second,  the relation of the present observations to the topological properties of the Kitaev chain has to be explored. 
Moreover, similar effects may be investigated in other topological models, as for example the $p_x+ip_y$ model,~\cite{PIP, TopSupRev} where the methods developed here can be directly applied. 
Third, the implementation of the optimal 
recovery operation should be also analyzed, because the required recovery time required  may need to  grow exponentially with the system size. 
Finally,  similar effects can be explored in other systems where the localization of information in Hilbert space is resilient to temperature and general interactions with the environment. Specifically, it is  intriguing to investigate the connection between our results and many-body  localization phenomena.~\cite{Altshuler,Huse}  Some of the concepts and tools developed in the present work can be used to explore the utility of such systems for storage and manipulation of quantum information.~\cite{ManyBodyHuse} 

\begin{acknowledgements}

We gratefully acknowledge fruitful discussions with G. Giedke, V. Giovannetti,  B. Horstmann, R. Huse, C. Kraus, F. Pastawski. Special thanks for invaluable support to H.-H. Tu.
Part of this work has been supported by the European
Community's Seventh Framework Programme (FP7/2007-2013) under grant agreement no. 247687 (IP-AQUTE). LM is supported by Regione Toscana POR FSE 2007-2013.

\end{acknowledgements}

\appendix

\section{Optimal Recovery Operations}

\subsection{General Recovery Operation}
\label{appendix1}

We present the derivation of Eq.~\eqref{eq:Fopt}.
We first show that the fidelity of any recovery operation is upper bounded by~\eqref{eq:Fopt}, $F(\mathcal R_t) \leq F^{\op}_t$ $\forall \mathcal R_t$,
and then construct an explicit recovery operation, ${\cal R}_t^{\op}$, which achieves the upper bound. 

We recall that for any bounded Hermitian operator $\hat X$ the trace norm is
 $\|\hat X\|_\tr= \max_{\hat H} \, \tr( \hat H \hat X)$,
 and the maximization is restricted to Hermitian operators fulfilling
 $ \| \hat H\|_{\rm op}\le 1,$
where $\| \hat X \|_{\rm op}$ denotes the maximum singular value of $\hat X$. 
Moreover, according to~\eqref{eq:F:average}, $F(\mathcal R_t)$ is  an average over the Bloch surface, which is best expressed as:~\cite{Bowdrey}
 \begin{equation}
 \label{eq:F2design}
 F(\mathcal R_t) = \frac{1}{2} + \frac{1}{12}\sum_{\alpha=x,y,z} {\rm tr}[\hat \sigma_\alpha {\cal T}_t (\hat \sigma_\alpha)].
 \end{equation}

We derive the upper bound. Because of the contractivity of the trace distance~\cite{NC} under $\mathcal R_t$, the inequality
$$\tr\left[\hat \sigma_\alpha \, {\cal T}_t(\hat \sigma_\alpha)\right]\le \|\hat \sigma_\alpha\|_{\rm op} \|{\cal T}_t(\hat \sigma_\alpha)\|_\tr \le 1 \cdot \|{\cal D}_t(\hat \sigma_\alpha)\|_\tr$$ 
holds, from which we obtain:
 \begin{equation}
 \label{eq:ApFopt}
 F(\mathcal R_t) \le \frac{1}{2}+\frac{1}{12} \sum_{\alpha=x,y,z} \| \hat \rho_{\alpha,+}(t)- \hat \rho_{\alpha,-}(t)\|_{\tr},
 \end{equation}
where $\hat \rho_{\alpha, \pm} (t) \doteqdot 
\mathcal D_t (\hat \Psi_{\alpha, \pm})$, with $\hat \Psi_{\alpha,\pm} \doteqdot (\hat {\mathbb I} \pm \hat \sigma_\alpha)/2$, $\alpha = x,y,z$. Notice that the states $\hat \rho_{\pm} (t)$ defined in Sec.~\ref{sec:rec:of:info} equal the $\hat \rho_{x,\pm}(t)$ which have been just defined.

Let us now specify~\eqref{eq:ApFopt} to the case of ${\cal D}_t$ not acting on the fermionic mode $\hat a$ (see Sec.~\ref{sec:setup}). 
The operator $\hat H_z \doteqdot \hat a \hat a^\dagger- \hat a^\dagger \hat a$ is Hermitian and $\| \hat H_z\|_{\rm op} \leq 1$; moreover
$\tr[\hat H_z\, \hat \rho_{z,\pm}(t)]= \pm 1$. 
Thus we have:
\begin{eqnarray}
2&=&\tr[\hat H_z(\hat \rho_{z,+}(t)-
\hat \rho_{z,-}(t))]\le \nonumber \\
&\le&
\|\hat \rho_{z,+}(t)-\hat \rho_{z,-}(t) \|_\tr\le  ||\hat \Psi_{z,+}-\hat \Psi_{z,-}||_\tr= 2. \nonumber
\end{eqnarray}
Thus: $\| \hat \rho_{z,+}(t)-\hat \rho_{z,-}(t)\|_\tr=2$.
Furthermore, since ${\cal D}_t$ does not act on $\hat a$, we can write 
$\hat \rho_{y,\pm}(t)=\hat V_{xy} \hat \rho_{x,\pm}(t) \hat V_{xy}^\dagger$, where $\hat V_{xy}\doteqdot e^{-i\pi \hat a^\dagger \hat a/2}$. Therefore:
 \begin{equation}
 \|\hat \rho_{x,+}(t)-\hat \rho_{x,-}(t)\|_\tr=
 \|\hat \rho_{y,+}(t)-\hat \rho_{y,-}(t)\|_\tr 
\label{eq:ugualexy}
 \end{equation}
The bound $F(\mathcal R_t) \leq F^{\op}_t$ $\forall \mathcal R_t$ follows from the combination of~\eqref{eq:ApFopt} with these considerations.

The recovery map $\mathcal R^\op_t$ which achieves $F(\mathcal R^\op_t) = F^{\op}_t$ is in the form: 
\begin{equation}
\label{Recov}
{\cal R}^{\op}_t(\hat \rho(t)) = 
\frac{1}{2} \hat {\mathbb I} \, \tr\left[\hat \rho (t) \right] + \frac{1}{2} \sum_\alpha \hat \sigma_\alpha \,\tr\left[\hat H_\alpha\hat \rho (t) \right].
\end{equation}
and the operators $\hat H_\alpha$ are such that $\tr[ \hat H_\alpha \mathcal D_t (\hat \sigma'_\alpha)] = \| \mathcal D_t (\hat \sigma'_\alpha)\|_\tr$. $\hat H_z$ has already been defined.
The $\hat H_{\alpha}$ can be interpreted as the observables to be measured in $\hat \rho(t)$ in order to reconstruct $\hat \rho_q$.

Let us rewrite 
$$
\hat \rho_{x,+}(t)-\hat \rho_{x,-}(t)=\hat a \hat R + \hat R^\dagger \hat a^\dagger; \quad
\hat R = \mathcal D_t 
\left(-\hat b \right). 
$$ 
We write the most general Hermitian operator: 
$$\hat H_x= \hat a \hat S_1+\hat S_1^\dagger \hat a^\dagger + \hat a^\dagger \hat a \hat S_{2}+ \hat a \hat a^\dagger \hat S_{3}$$ which must satisfy $\hat H_x^\dagger \hat H_x\le \hat {\mathbb I}$. We get 
$$
\tr\left[\hat H_x \left(\hat a \hat R + \hat R^\dagger \hat a^\dagger \right)\right]=\tr\left[
\left(\hat S_1 \hat R^\dagger+\hat R \hat S_1^\dagger \right) \hat a^\dagger \hat a\right]
$$ 
Using the left polar decomposition $\hat R=\hat P \hat U$, where $\hat P=\sqrt{\hat R \hat R^\dagger}$ is positive semi-definite and $\hat U$ is unitary, we have that the maximum is attained when $\hat S_1=\hat U$, and $\hat S_{2}=\hat S_3=0$. Therefore, 
the  maximum is achieved by:
$$
\hat H_x=\hat a \hat U+ \hat U^\dagger \hat a^\dagger; \qquad \hat H_y=-i \hat a \hat U+i \hat U^\dagger \hat a^\dagger.
$$ 
Furthermore, since both $\hat a$ and $\hat R$ change the fermionic parity, $\hat H_{x,y}$ do not. 
They also fulfill:
\begin{equation}
\tr \left[\hat H_\alpha \hat \rho_{\alpha,\pm}(t)\right] = \pm
\frac{1}{2}\|\hat \rho_{x,+}(t)-\hat \rho_{x,-}(t)\|_\tr; \; \; \; \alpha=x,y \nonumber
\end{equation}
from which $\tr[ \hat H_\alpha \mathcal D_t (\hat \sigma_\alpha)] = \| \mathcal D_t (\hat \sigma_\alpha)\|_\tr$ follows.

The optimal recovery map~\eqref{Recov}
is linear, trace preserving, and it also preserves the fermionic parity, since $\hat H_\alpha$ do. For it to be a valid quantum channel we have only to show that it is completely positive. We construct a unitary operator acting on all the fermionic modes, $\hat W$, such that ${\cal R}^{\op}_t(\hat \rho)= 
\tr\left[\hat W \hat \rho \hat W^\dagger\right]$, where the trace is taken over the fermionic degrees of freedom which are different from $\hat a$ and $\hat b$ and the qubit is recovered in the even parity sector of their Hilbert space. 
The operator is:
 \begin{equation}
 \hat W=\frac{1}{8} \left[\hat 1 + \sum_{\alpha=x,y,z} \hat \sigma'_\alpha \hat H_\alpha\right].
 \end{equation}
Using that $\hat H_\alpha^2=\hat 1$ and $\hat H_\alpha \hat H_\beta = i \epsilon_{\alpha,\beta,\gamma} \hat H_\gamma$, where $\epsilon_{\alpha,\beta,\gamma}$ is the Levi-Civita symbol, 
one can show that $\hat W$ is unitary and that it correctly defines
${\cal R}^{\op}_t(\hat \rho) $
Furthermore,  $F({\cal R}_t^{\op})$ saturates the bound in~\eqref{eq:Fopt}.

\subsection{Gaussian Recovery Operation}
\label{appendix2}

We present the derivation of Eq.~\eqref{eq:FGopt};  
the proof is similar to that in Appendix~\ref{appendix1}.
Because Gaussian fermionic channels map fermions to fermions, we need to explicitly consider the fact that the recovered qubit is composed of fermions. 
In particular, we have to consider what happens when the decoherence channel changes the parity of the state but still preserves the orthogonality of the initial qubit states.
This may occur, for instance, when decoherence is caused by the coherent interchange of particles with a reservoir. 
Imagine the initial qubit state $\ket \Phi = \gamma \ket 0 + \delta \ket 1$,  $|\gamma|^2 + |\delta|^2 = 1$, is mapped to (see definitions in Sec.~\ref{sec:setup}): 
$$
\ket {\Phi (t)} = 
\alpha \left( \gamma \ket 0 + \delta \ket 1  \right)
+
\beta \left( \gamma \ket{ \tilde 0} + \delta \ket{ \tilde 1 }  \right);
\;
|\alpha|^2 + |\beta|^2 = 1
$$
The state is non Gaussian (it is pure but with no defined parity) and clearly contains all the information about the initial state. However, no Gaussian recovery operation can bring it back to $\ket \Phi$, which is a Gaussian state.
The problem can be bypassed defining the recovered (fermionic) qubit in both parity sectors.

Recalling that $\hat m_1$, $\hat m_2$ are the decoherence-free modes which constitute $\hat a$ and that the zero-energy modes of the second Kitaev chain are $\hat m_3$ and $\hat m_4$ (see Sec.~\ref{sec:setup}), we express the Pauli operators of the qubit as:
 \begin{eqnarray}
 \hat \sigma''_x&=& (\hat a^\dagger - \hat a)(\hat b^\dagger + \hat b)=i \hat m_3 \hat m_2;\\
 \hat \sigma''_y&=&-i(\hat a^\dagger +\hat a)(\hat b^\dagger + \hat b)=i \hat m_3 \hat m_1;\\
 \hat \sigma''_z&=& \hat a^\dagger \hat a- \hat a \hat a^\dagger=i\hat m_1 \hat m_2.
 \label{eq:sigmas}
 \end{eqnarray}
 where the modes $\hat a^{(\dagger)}$ and $\hat b^{(\dagger)}$ have been defined in Sec.~\ref{sec:setup}.
Notice the difference with the $\hat \sigma'_{\alpha}$ in Eq.~\eqref{eq:paulitilde}: the $\hat \sigma''_\alpha$ act on both the parity sectors of the qubit.

We recall that the action of a general Gaussian channel transforms a Gaussian $M$-modes state with CM: $\Gamma$ into a $N$-modes state with CM: $\Gamma'=B \Gamma B^T + A$ (see Appendix~\ref{sec:FGstates} and Ref.~[\onlinecite{FGS}]). 
$B$ and $A$ are  $2N\times 2M$ and $2N\times 2N$ matrices chosen such that $Q$, defined as:
\begin{equation}
Q = \left(
\begin{array}{cc} A & B \\ -B^T & 0 \end{array}
\right)
\label{eq:M:channel}
\end{equation}
satisfies $Q^T Q \leq \mathbb I$; $A$ must be skew-symmetric.

We denote $\mathcal R_{\mathrm G,t}$ the Gaussian recovery operation, and $\mathcal T_{\mathrm G, t} = \mathcal R_{\mathrm G,t} \circ \mathcal D_t$.
Moreover, we define:
 \begin{eqnarray}
\Delta_\alpha &=& \Gamma_{{\cal D}_t(\hat \Psi_{\alpha,+})} - \Gamma_{{\cal D}_t (\hat \Psi_{\alpha,-})}, \\
\Delta^{\rm out}_\alpha &=& \Gamma_{{\cal T}_{\mathrm G,t}(\hat \Psi_{\alpha,+})} - \Gamma_{{\cal T}_{\mathrm G, t}(\hat \Psi_{\alpha,-})} = B_{\cal R} \Delta_{\alpha} B_{\cal R}^T, \quad
 \end{eqnarray}
the difference of covariance matrices corresponding to the states after the decoherence channel and after the recovery operation, respectively. 
Notice that the matrices $\Delta_\alpha$ are $2N \times 2N$ matrices, whereas the matrices $\Delta^{\rm out}_{\alpha}$ are $4 \times 4$ matrices.
The assumption that the decoherence channel does not act on the first two Majorana modes $\hat m_{1,2}$ is reflected by some properties of $\Delta_\alpha$ which are best expressed considering
the block structure:
 \begin{equation}
 \Delta_{\alpha} = \left(
 \begin{matrix}
 K_\alpha'    & -L^T_\alpha \\
 L_\alpha & K_\alpha''
 \end{matrix}
 \right),
 \end{equation}
where $K',L$ and $K''$ are $2\times 2$, $2(N-1)\times 2$, and $2(N-1)\times 2(N-1)$ matrices, respectively. 
We obtain:
 \begin{equation}
 K_z' = -2 J; \; \;
 L_x=\left( \begin{matrix} \vec l_1 ,& \vec l_2 \end{matrix}  \right);
\;\;
 L_y = L_x J;
\;\; 
J= \left( \begin{array}{cc} 0 & -1 \\ 1 & 0 \end{array} \right);
\label{eq:def:KLL}
 \end{equation}
where $\vec l_{1,2}$ are column vectors. 
Additionally, $L_z$, $K_{x,y}'$ and $K_{x,y}''$ are zero matrices. 
Thus:
\begin{equation}
\label{eq:Delta}
\|\Delta_x\|_{\rm op} = \|\Delta_y\|_{\rm op} = 
\| L_x \|_{\rm op}  \leq 2.
\end{equation}

We can now show that $F(\mathcal R_{\mathrm G,t}) \leq F^\op_{\mathrm G,t}$ $\forall \mathcal R_{\mathrm G,t}$. 
The starting point is equation~\eqref{eq:F2design}, modified as follows:
\begin{equation}
 F_{\mathrm G,t} = \frac{1}{2} + \frac{1}{12}\sum_{\alpha=x,y,z} {\rm tr}[\hat \sigma''_\alpha {\cal T}_{\mathrm G,t} (\hat \sigma_\alpha)].
 \label{eq:largebound}
\end{equation}
Noting that the $\hat \sigma''_\alpha$ are quadratic in the Majorana operators and recalling the definition of CM, we obtain:
\begin{equation}
\tr\left[\hat \sigma''_\alpha \mathcal T_{\mathrm G,t}( \hat \sigma_\alpha)\right] = \left(\Delta^{\rm out}_\alpha\right)_{\beta_1,\beta_2}
\le \|\Delta^{\rm out}_\alpha\|_{\rm op},
\end{equation}
where $(\beta_{1}, \beta_2)=(3,2)$, $(3,1)$, and $(1,2)$ for $\alpha=x,y,z$, respectively. 
The most general Gaussian recovery operator $\mathcal R_{\mathrm G, t}$ yields: 
\begin{equation}
\|\Delta^{\rm out}_\alpha\|_{\rm op}=\|B_{\cal R} \, \Delta_\alpha \, B_{\cal R}^T\|_{\rm op} \le \|\Delta_\alpha\|_{\rm op}
\end{equation}
given that $Q^TQ\le \mathbb I$. 
Since $2\ge \|\Delta_z \|_{\rm op} \ge (\Delta_z)_{1,2}=(K_z)_{1,2}=2$, we get that $\|\Delta_z\|_{\rm op}=2$. Using (\ref{eq:Delta}) we obtain the desired bound.

We now provide an explicit Gaussian recovery operation which attains the bound. 
The recovery operation consists of the application of a Gaussian unitary operation $\hat W_{\rm G}$ to the the system and in subsequently tracing out $N-2$ modes of the system.
To define $\hat W_{\rm G}$, consider the singular value decomposition of $L_x = U \Sigma V^T$, where $U$ ($V$) is a unitary $2(N-1)\times 2(N-1)$ ($2 \times 2$) matrix and $\Sigma$ is a $2(N-1)\times 2$ matrix. Clearly, it is also possible to construct $U'$ and $V'$ such that $L_x = U' \Sigma' V'$ and $\Sigma'$ has at most two elements different from zero, $\Sigma_{1,2} \geq \Sigma_{2,1} \geq 0$, named the singular values of $L_x$.
We define $\hat W_{\rm G}$ to be the unitary transformation which is represented by an orthogonal transformation $V' \oplus U'$:
\begin{equation}
\hat W_{\rm G} \, \vec {\hat c} \, \hat W_{\rm G}^\dagger = \left( V' \oplus U'\right) \; \vec {\hat c}
\end{equation}
Physically, $\hat W_{\rm G}$ rotates all the information between ancilla and system into the first two modes of the system. The other modes can be now traced out. Summarizing:
\begin{equation}
\mathcal R^{\op}_{\mathrm G, t} (\hat \rho) = \tr \left[ 
\hat W_{\rm G} \, \hat \rho \, \hat W_{\rm G}^\dagger
\right]
\end{equation}
The CM of $\mathcal R^{\op}_{\mathrm G, t} (\hat \rho)$ is a function of $\Gamma$, the CM of $\hat \rho$:
\begin{equation}
 \Gamma_{\mathcal R^{\op}_{\mathrm G, t} (\hat \rho)} = \left.
 \left[ 
 (V' \oplus U') \, \Gamma \, (V'^T \oplus U'^T)
 \right] \right|_{(1-4),(1-4)}.
\end{equation}

Finally, let us prove that $F(\mathcal R^{\op}_{\mathrm G,t}) $ saturates the bound. Denote $\mathcal T^{\op}_{\mathrm G, t} = \mathcal R^{\op}_{\mathrm G, t} \circ \mathcal D_t $.
Clearly:
\begin{equation}
\sum_{\alpha=x,y,z} \hspace{-0.2cm} \tr\left[\hat \sigma''_\alpha \mathcal T^{\op}_{\mathrm G,t}(\hat \sigma_\alpha)\right] = 
\left(\Delta^{\rm out}_x\right)_{3,2} +
\left(\Delta^{\rm out}_y\right)_{3,1} +
\left(\Delta^{\rm out}_z\right)_{1,2}.\nonumber
\end{equation}
By construction of $\hat W_{\rm G}$, $\left(\Delta_x^{\rm out} \right)_{3,2} =  \| \Delta_x \|_{\rm op}$. Since $J$ commutes with every $2\times 2$ orthogonal matrix, $\left(\Delta_y^{\rm out} \right)_{3,1} =  \| \Delta_y \|_{\rm op} = \| \Delta_x \|_{\rm op}$. Finally, $\left(\Delta^{\rm out}_z\right)_{1,2} = \left(\Delta_{z}\right)_{1,2}=2$ because the orthogonal transformation $V'$ leaves the covariance matrix of the fermionic mode $\hat a$ unchanged. Together with equation~\eqref{eq:largebound}, this shows that the recovery operation $\mathcal R^{\rm opt}_{\mathrm G, t}$ saturates the bound in~\eqref{eq:FGopt}.

\section{Fermionic Gaussian States}
\label{sec:FGstates}

The formalism of fermionic Gaussian states~(FGS)~\cite{FGS} is particularly useful
in the treatment of quadratic fermionic theories, which include a wide class of topological materials.~\cite{TopSupRev}
Not only FGS comprise the ground states and thermal states of such Hamiltonians  (via Bogoliubov transform)
but also they describe dynamical evolution under some master equations (exactly)~\cite{FGDissipative}
or in presence of moderate interactions (approximately).~\cite{Kraus}
For a $N$ modes system, calculations are restricted on a space scaling only as $N$ rather than exponentially.
Indeed, these states are fully characterized by the sole (antisymmetric and real valued) covariance matrix (CM),
i.e. the expectation values of quadratic combinations of fields, whereas all higher moments can be deduced
via Wick theorem.~\cite{TDV}

Given $N$ fermionic modes, we can conveniently rewrite the $2N$ canonical Dirac fermionic operators
$\{\hat d_i^{(\dagger)}\}_{i=1 \ldots N}$ (with $\{ \hat d_i, \hat d_j\} = 0;  \ \{\hat d_i,\hat d^{\dagger}_j\} = \delta_{i,j}$)
in terms of Majorana operators, i.e. fermionic operators which are real, Hermitian and unitary:
\begin{equation}
\small{
\hat c_{2j-1} = \hat d_j + \hat d^{\dagger}_j; \quad
\hat c_{2j} = -i (\hat d_j - \hat d^{\dagger}_j); \quad
\{ \hat c_{m}, \hat c_{n}\} = 2 \delta_{m,n},
}
\label{eq:Majo}
\end{equation}
where $m \in \{(1,1), \ldots (N,2) \}$ glues two sub-indices together for brevity.
Some simple algebra shows that Eq.~\eqref{eq:protecting} is a rewriting of 
\begin{equation*}
\small{
 \hat H(\mu) = \sum_{j=1}^{N-1}  \left( - J \hat d_{j}^\dagger \hat d_{j+1}^{\phantom{\dagger}}
 +\Delta \hat d_{j}^\dagger \hat d_{j+1}^{\dagger} + \mathrm{h.c.} \right)
 - \mu \sum_{j=1}^{N} \hat d_{j}^\dagger \hat d_{j}^{\phantom{\dagger}}.
}
\label{eq:Hmu}
\end{equation*}
Canonical transformations can be represented by orthogonal real matrices,
$\hat c_k \to \hat c'_k = \sum_l O_{k,l} \hat c_l$, as well as by a unitary rotation $\hat c'_k = \hat U \hat c_k \hat U^\dagger$
in Fock space:~\cite{FGS} in the case of $O \in O(2m)$, i.e. $\det O =1$, the relation reads
$O = \exp(A)  \Leftrightarrow \hat{U} =  \exp{(-A_{\alpha,\beta} \hat{c}_\alpha \hat{c}_\beta /4)}$ apart from an arbitrary phase.
The number parity operator reads $\hat P = (-1)^{\sum_j \hat d^\dagger_j \hat d_j} = i^N \prod_k \hat c_k$
and is almost invariant under canonical transformations: $\hat P' = \det O \cdot \hat P$.

A $N$-modes FGS is a $N$-fermions state which has a density operator of the form
$\hat{\rho} = \prod_{\alpha=1}^{N} \hat{\rho}_\alpha$, with
\begin{equation}
\small{
\hat{\rho}_{\alpha} = \frac{e^{- \beta_\alpha \hat d_{\alpha}^{\dagger} \hat d_{\alpha}^{\phantom{\dagger}}}}{1+e^{-\beta_\alpha}} =
 \frac{e^{- \frac{i}{4} \beta_\alpha ( \hat c_{\alpha,1} \hat c_{\alpha,2} - \hat c_{\alpha,2} \hat c_{\alpha,1})}}{2 \cosh (\beta_\alpha/2)} =
\frac{\hat 1 - i \lambda_\alpha \hat c_{\alpha,1} \hat c_{\alpha,2}}{2},
}
\label{eq:rho}
\end{equation}
where $\lambda_\alpha = \tanh (\beta_\alpha / 2)$, and the $\hat a_{\alpha}^{(\dagger)}$ ($\hat c_{\alpha,\sigma}$)
are the eigenmodes of the density operator.
One can easily verify that $\text{Tr} \hat{\rho} = 1$, whereas $\text{Tr} \hat{\rho}^2 = \prod_\alpha (1+\lambda_\alpha^2)/2$,
i.e. the state is pure if and only if $\lambda_\alpha = \pm 1$; 
moreover,  $\hat{\rho}$ is positive -- and thus a well-defined density operator -- if and only if $\lambda_\alpha \in [-1,1]$.
FGS automatically satisfy the superselection rule of the fermionic parity $\hat P$,
and therefore their density matrices can be expressed as a direct sum $\hat \rho = \hat \rho_{\rm even} \oplus \hat \rho_{\rm odd}$.

The skew-symmetric CM of a FGS $\hat{\rho}$ is defined as
the table of quadratic expectation values:
\begin{equation}
\small{
\Gamma_{m,n} = \frac{i}{2} \mathrm{Tr} \left[
\hat{\rho} \, (\hat c_{m} \hat c_{n} - \hat c_{n} \hat c_{m})
\right];
\ \ 
\Gamma =
\bigoplus_{\alpha}
\left(
\begin{array}{cc}
0 & -\lambda_{\alpha} \\ \lambda_{\alpha} & 0
\end{array}
\right),
}
\label{eq:gamma}
\end{equation}
where the second expression is given in the eigenbasis of Eq.~\eqref{eq:rho}.
Under a canonical transformation $O$, the CM transforms as $\Gamma' = O \Gamma O^T$.
The CM completely characterizes the properties of a FGS, as  elegantly stated by the following reformulation
of the Wick's theorem:~\cite{TDV}
\begin{equation}
i^p \;\text{Tr}[\hat \rho \, \hat c_{\alpha_1} \ldots \hat c_{\alpha_{2p}}] = 
\mathrm{Pf} \left[\left. \Gamma \right|_{\alpha_1 \ldots \alpha_{2p}} \right],
\label{eq:wick}
\end{equation}
where $\left. \Gamma \right|_{\alpha_1 \ldots \alpha_{2p}} $ is the restriction of $\Gamma$ to the modes $\{\alpha_1 \ldots \alpha_{2p}\}$,
and $\mathrm{Pf}$ denotes its Pfaffian.
This allows to simulate FGS efficiently with classical computers.

The squared overlap of two FGS $\hat{\rho}$ and $\hat{\sigma}$ is:~\cite{FGS}
\begin{equation}
\mathrm{Tr} \left[ \hat{\rho} \, \hat{\sigma} \right] = + \sqrt{
\det
\left[ \frac{1 - \Gamma_{\rho}\Gamma_{\sigma}}{2} \right]
} \ ,
\label{eq:overlap}
\end{equation}
Moreover, generalizations of formula~\ref{eq:overlap} can be derived for any two Gaussian operators.~\cite{FGS} 
Because the time-evolution $\hat U(t)$ under a quadratic Hamiltonian $\hat H$ is a Gaussian operator,  such formulas for $\text{Tr} [ \hat{\rho} \, \hat{U}(t) ]$ and $\text{Tr} [ \hat{U}'(t) \, \hat{U}''(t) ]$ have been widely used in the main text. Moreover, the CM of $\hat U(t) \hat \rho(0) \hat U(t)^\dagger$ is: $\Gamma(t) = O(t) \Gamma(0) O(t)^T$, where $O(t)= e^{- T t}$ and $T$ is the real skew-symmetric matrix such that $\hat H = \frac i4 \sum_{j,k} T_{j,k} \hat c_{j} \hat c_k$.

Also the Uhlmann fidelity among mixed Gaussian states, $F_U(\hat \rho,\hat \sigma) = \text{Tr} \sqrt{\hat \rho^{1/2} \hat \sigma \hat \rho^{1/2}}$,
can be efficiently computed via their CM's. 
By using Eq.(\ref{eq:rho}), we can define $\hat H_{\rho}$ such that
$\hat \rho^{1/2} = \exp(- \hat H_{\rho}) / \sqrt{\mathrm{Tr} \exp(- 2 \hat H_{\rho})}$
and the corresponding imaginary-time evolution of the state $\hat \sigma$:~\cite{Kraus}
\begin{equation}
\small{
\hat \rho^{1/2} \,\hat \sigma \,\hat \rho^{1/2} = \mathrm{Tr} [ \hat \rho \, \hat \sigma] \cdot \hat{\sigma}_I(\tau = 1) ;
\quad
\hat{\sigma}_I(\tau) = \frac{e^{-\hat H_{\rho} \tau} \hat \sigma e^{-\hat H_{\rho} \tau}}{\text{Tr} \left[e^{-2 \hat H_{\rho} \tau} \hat \sigma\right]} \ .
}
\end{equation}
Since $\hat{\sigma}_I(\tau)$ is still a Gaussian state, whose CM can be efficiently computed,~\cite{Kraus}
the trace of its square root in $F_U$ can be calculated by looking again at Eq.~(\ref{eq:rho}) as above.

\section{Some Details on Hamiltonian Perturbations}\label{app:Ham:pert}

\begin{figure}[t]
\begin{center}
\includegraphics[width=\columnwidth]{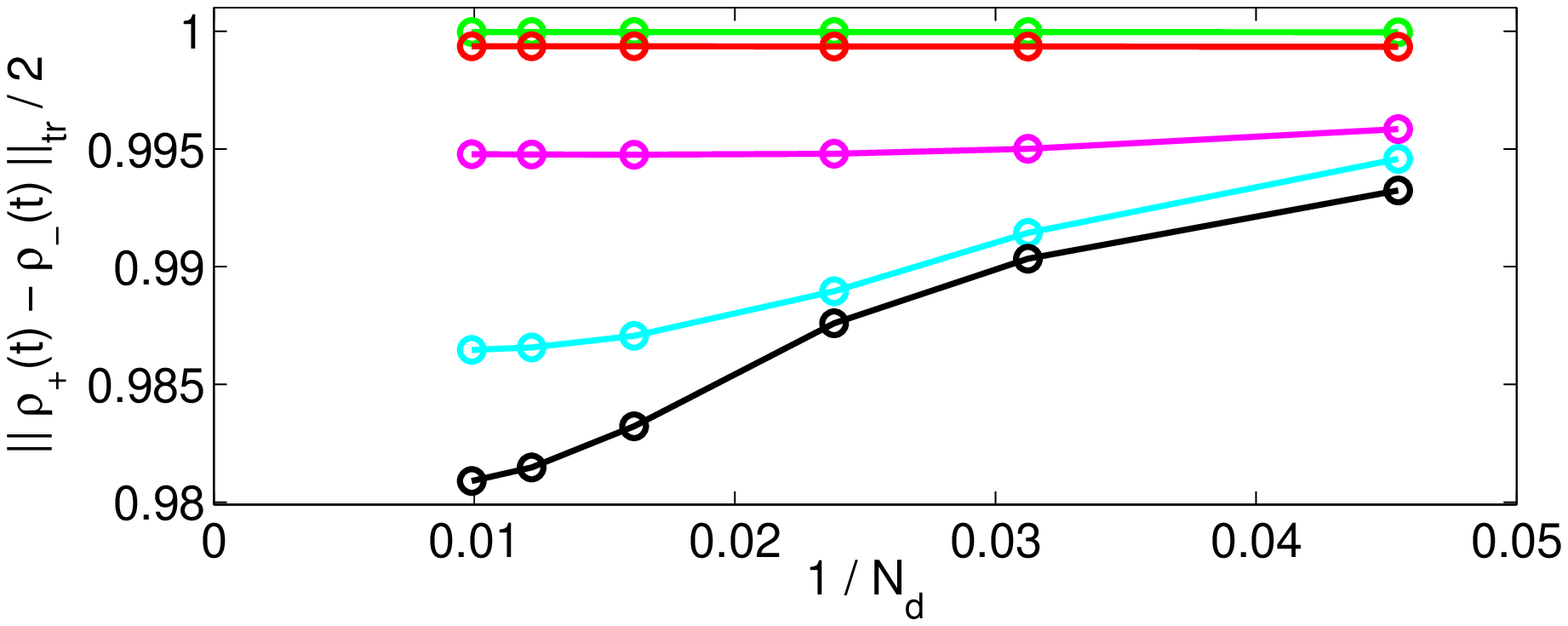}
\caption{Dependence of $F_t^\opt$ on $N_{\rm d}$. These data refer to the topological perturbation of Fig.~\ref{fig:decomap:unknownHam}: $N=32$, $\mu_0 = 0$, $\mu_-=J$, $\mu_+=1.5J$. The scaling is shown for five times. From up to down: $24 J^{-1}$, $54 J^{-1}$, $174 J^{-1}$, $354 J^{-1}$, $474 J^{-1}$. Data show a convergence behaviour for $N_{\rm d} \to \infty$.}
\label{Fig:1}

\includegraphics[width=\columnwidth]{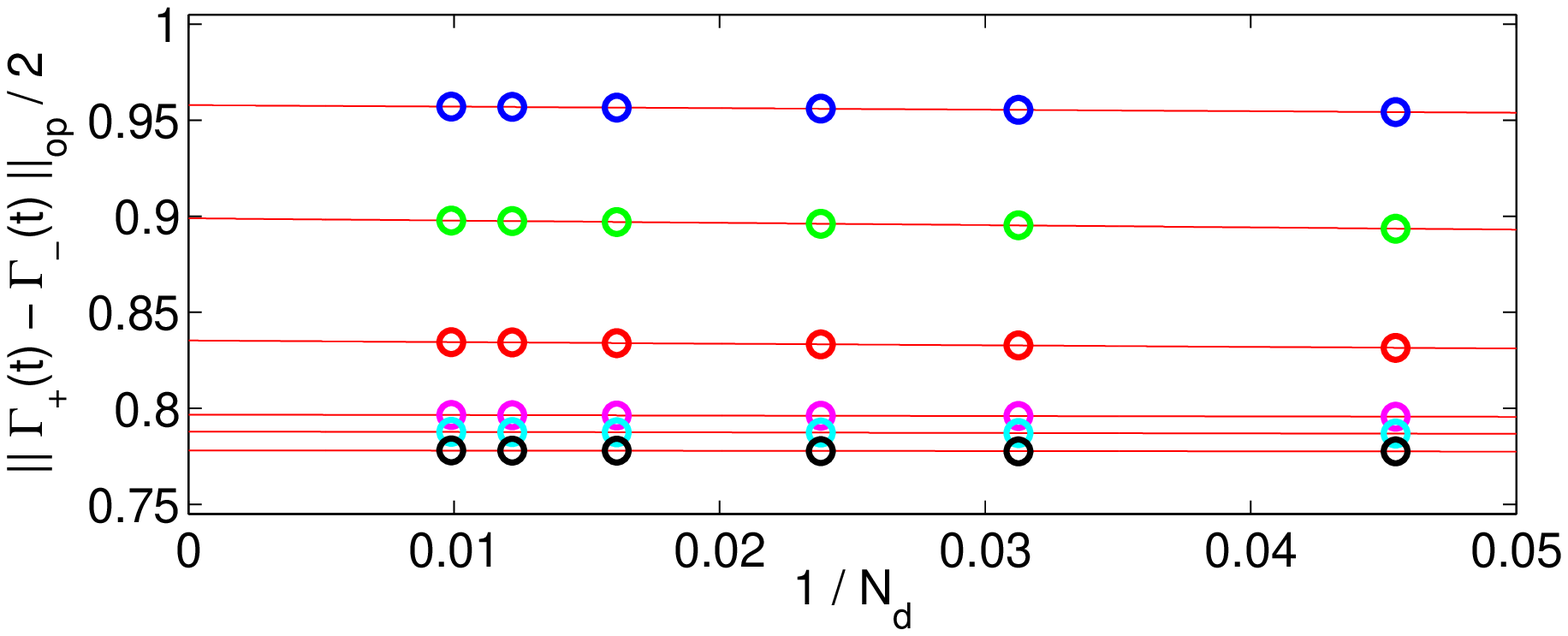}
\caption{Dependence of $F_{\mathrm G t}^\opt$ on $1 / N_{\rm d}$. These data refer to the topological perturbation of Fig.~\ref{fig:decomap:unknownHam}: $N=32$, $\mu_0 = 0$, $\mu_-=J$, $\mu_+=1.5J$. The scaling is shown for five times. From up to down: $12 J^{-1}$, $24 J^{-1}$, $54 J^{-1}$, $174 J^{-1}$, $354 J^{-1}$, $474 J^{-1}$. Data show a $N_{\rm d}^{-1}$ dependence: the $N_{\rm d} \to \infty$ value is extrapolated with a linear fit (thin red lines).}
\label{Fig:2}

\end{center}
\end{figure}

We present a demonstration of Eq.~\eqref{eq:thesis1}.
First, a simple inspection shows that $G_\sigma = G_\sigma^\dagger$, $G_\sigma >0$ and  $\| \sqrt{G_{\sigma}/N_{\rm d}}\|_{\rm HS}=1$.
Let us consider the states $\ket{\pm} = 
\left( \ket{ 0} \pm \ket{1} \right)/\sqrt 2$
and use them to define the overlap matrix $M$:
\begin{equation}
 M =  \left(
\begin{array}{cc} M^{(+,+)} & M^{(+,-)} \\
M^{(-,+)} & M^{(-,-)} \end{array} \right)= \frac 12 \left(
\begin{array}{cc} G_0+G_1 & G_0-G_1 \\G_0-G_1 & G_0+G_1 \end{array} \right).
\end{equation}
with $M^{(\tau,\tau')}_{j,k} = \langle \tau | e^{i \hat H_j t} e^{-i \hat H_k t} \ket{\tau'}$ ($\tau, \tau' = \pm$). The second equality follows from
the assumption of a decoherence-free mode.
Let us now consider the set of $2 N_{\rm d}$ states: 
$\mathcal C = \{ e^{-i \hat H_j t} \ket{ +}\}_{j=1}^{N_{\rm d}}  \cup \{
e^{-i \hat H_j t} \ket{ -}\}_{j=1}^{N_{\rm d}}$ and an orthonormal basis $\mathcal B = \{\ket{x_j} \}_{j=1}^{N_B}$ such that: $\text{span } \mathcal B \equiv \text{span } \mathcal C$.
A matrix $Y$ representing the basis change: $e^{-i \hat  H_k t} \ket{+} = \sum_q Y^*_{k ,q} \ket{x_q}$ and $e^{-i \hat H_k t} \ket{-} = \sum_q Y^*_{N_{\rm d}+k ,q} \ket{x_q}$ is defined by $Y Y^\dagger=M$ and can be computed via the eigenvalue decomposition of $M$.
Given $V$ unitary and $D$ diagonal matrix such that: $M = V D V^\dagger$, a $Y$ defined as $Y \doteqdot V \sqrt D$ is one such possible basis change.
A simple algebra leads to:
\begin{equation}
\frac 12 \left\| \hat \rho_{+}(t) - \hat \rho_{-}(t)\right\|_\tr =
\frac 1 {2N_{\rm d}} \left\|
Y^\dagger \left( 
\begin{array}{cc} \mathbb I & 0 \\ 0 & - \mathbb I \end{array}
\right) Y
\right\|_\tr
\end{equation}
We can obtain a more explicit expression of $Y$ considering the eigenvalue decomposition of $G_\sigma$:
$
G_\sigma = V_\sigma D_\sigma V_\sigma^\dagger
$
and observing that the matrix:
\begin{equation}
 V = \left( 
\begin{array}{cc} V_0 & V_1 \\ V_0 & -V_1 \end{array}
\right)
\end{equation}
diagonalizes $M$. Some simple algebra shows that
the singular values of $Y^\dagger \left( 
\begin{array}{cc} \mathbb I & 0 \\ 0 & - \mathbb I \end{array}
\right) Y$ are two-fold degenerate and coincide with the singular values of $\sqrt{G_0 G_1}$. Equation~\eqref{eq:thesis1} follows from the positivity of $G_\sigma$.

The numerical computation of $G_\sigma$ is efficient when $\hat H_j$ is a quadratic Hamiltonian so that $\hat U_{j,k} \doteqdot e^{i  \hat H_j t} e^{-i  \hat H_k t}$ is a Gaussian operator ($\hat \rho_\sigma =\ket{g_\sigma} \bra{g_\sigma}$ is FGS) and thus $[G_{\sigma}]_{j,k}$ is a function of the CM of $\hat \rho_\sigma$ and $\hat U_{j,k}$. We warn the reader interested in reproducing the data that some care is required in order to obtain the proper phase.

We checked the possibility of using the data to obtain information regarding the limit $N_{\rm d} \to \infty$.
In Fig.~\ref{Fig:1} we show the dependence of  $F_t^\op$ for different values of $N_{\rm d}$; data show a clear convergence behaviour, even if the functional form of such scaling was not found. In figure~\ref{Fig:2} we show the dependence of $F_{\mathrm G, t}^\op$ on $N_{\rm d}$; data show a clear $\/N_{\rm d}$ scaling, which allowed us to take the limit $N_{\rm d} \to \infty$ with a linear fit.

\begin{figure}[t]
\begin{center}
\includegraphics[width=0.49\columnwidth]{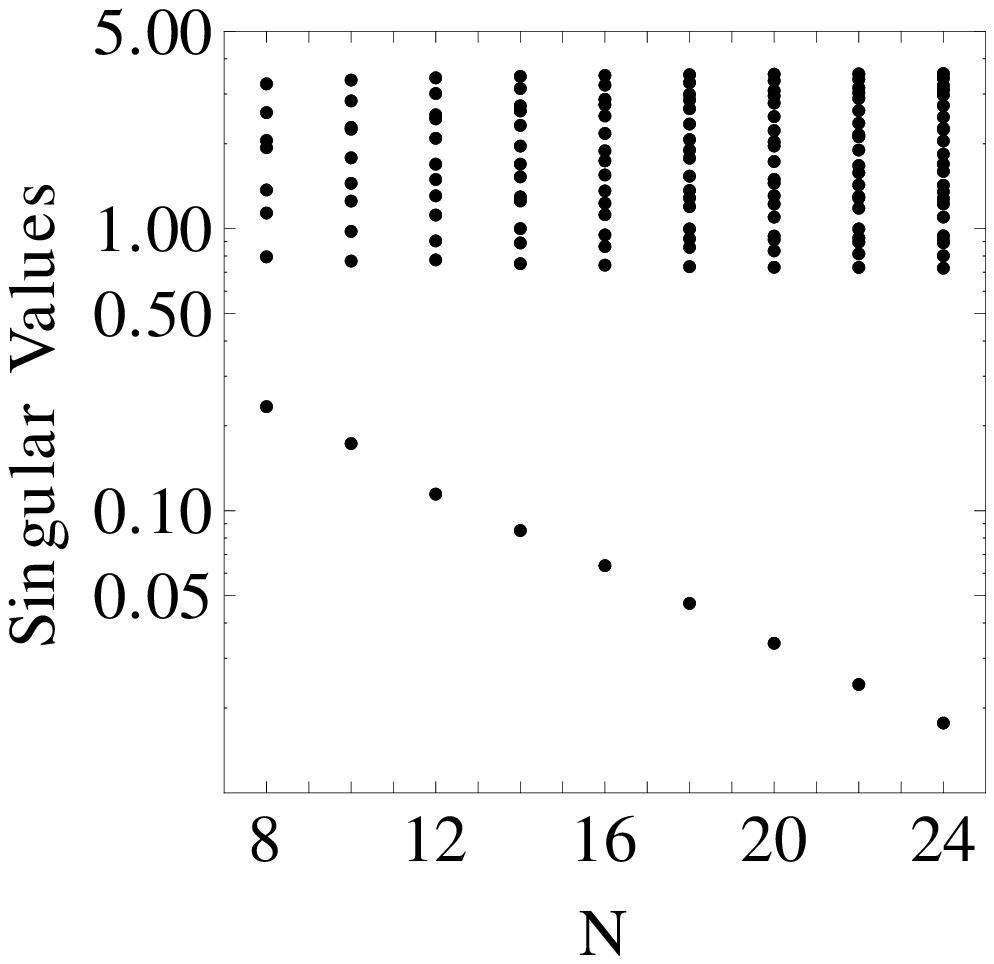}
\includegraphics[width=0.49\columnwidth]{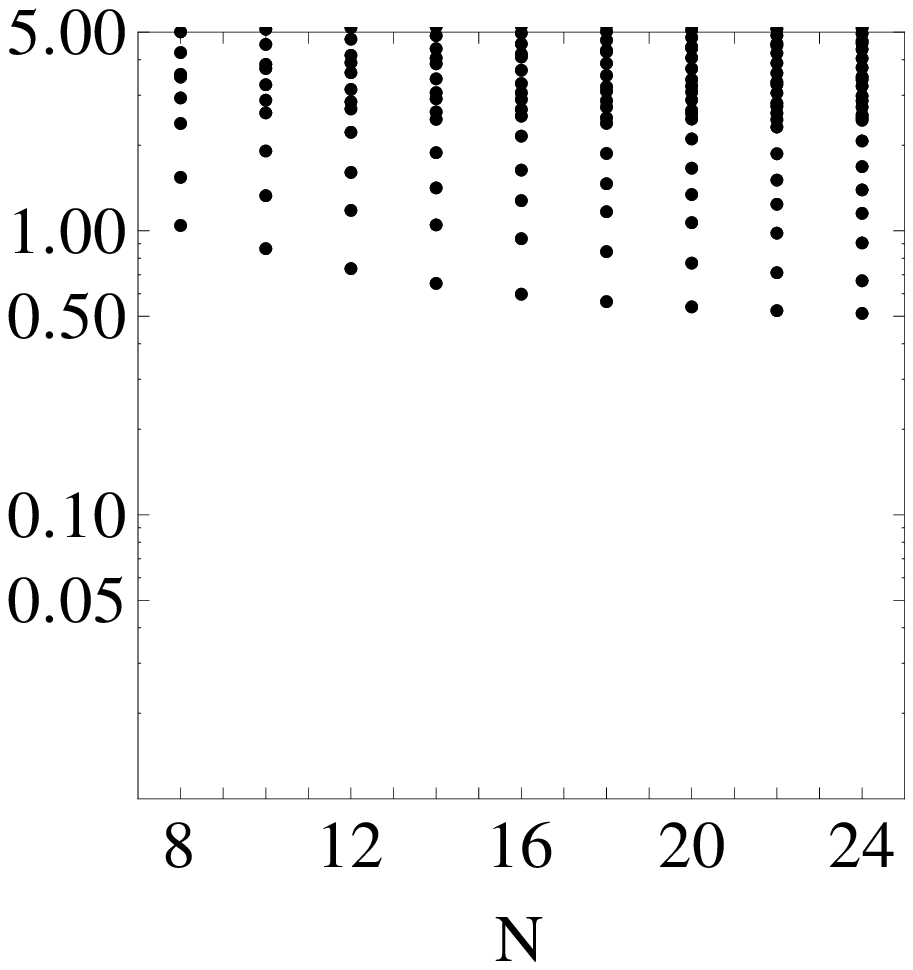}
\caption{Singular values of the matrix in equation~\eqref{eq:strangematrix}. (Left) The quench Hamiltonians are in the topological phase: $\mu_0 = 0$, $\mu_j = J$, $\mu_k = 1.5 J$. (Right) The quench Hamiltonians are not in the topological phase: $\mu_0 = 0$, $\mu_j = 2.5 J$ and $\mu_k = 3.0 J$.}
\label{figure3}
\end{center}
\end{figure}

In the main text we provide numerical evidence that the memory time of the system increases exponentially while letting the system size $N \to \infty$, which is equivalent to $X \doteqdot G_0 - G_1 \xrightarrow{N \to \infty} 0$. 
We first observe that:
\begin{equation}
  \left| X_{j,k} \right| \propto
 \left| 
\pf \left[\frac{\Gamma_{0}+\Gamma_{1}}{2}+\frac{\Upsilon}{\tr[\hat U_{j,k}(t)]} \right] \right|;
\label{eq:strangematrix}
\end{equation}
where $\Gamma_\sigma$ is the CM of $\hat \rho_\sigma$ and $\Upsilon$ is the CM of $\hat U_{j,k}$; we assume $\tr[\hat U_{j,k}(t)] \neq 0$ in order to avoid lengthy regularized expressions.
The proportionality factor between l.h.s. and r.h.s. is  bounded by $1$.
Direct numerical inspection of the matrices shows that
$X_{j,k} \xrightarrow{N \rightarrow \infty} e^{-N}$ because $\frac12 (\Gamma_{0}+\Gamma_{1})+\Upsilon/\tr[\hat U_{j,k}(t)]$ has two singular values which scale exponentially to zero~(see figure~\ref{figure3}). This is clear when $\hat U_{j,k} = \mathbb I $ and $\Upsilon = 0$ since  $\Gamma_{0}+\Gamma_{1}$ has two zero eigenvalues corresponding to the zero-energy modes of the Hamiltonian $H_0$. When $\hat U_{j,k} \neq \mathbb I $ it would be tempting to interpret $\Upsilon/\tr[\hat U_{j,k}(t)]$ as a perturbation and invoke some topological stability argument; unfortunately  $\Upsilon/\tr[\hat U_{j,k}(t)]$ is neither bounded nor it is short-range. Moreover, one would like to have an explanation which distinguishes the situation in which $\hat U_{j,k}(t)$ is the product of two time-evolution according to topological and non-topological Hamiltonians.
Intuitively, the argument must reside on the fact that any topological Hamiltonian spreads the localized zero-energy modes exponentially slower than non-topological Hamiltonians do.

\bigskip

In the main text we present results also for thermal states. Because in this case the matrix $\hat \rho_\pm(t)$ is not a convex combination of a limited number of known pure states, we cannot exactly compute $F^\opt_t$. Using the fact that $\hat \rho_{\pm} (t) = \frac 1{N_{\rm d}} \sum_{j=1}^{N_{\rm d}} e^{-i \hat H_j t} \hat \rho_{\pm}(0) e^{i \hat H_j t}$ and denoting 
$\hat \rho^{(j)}_{\pm} (t) = e^{-i \hat H_j t} \hat \rho_{\pm}(0) e^{i \hat H_j t} $, we compute the following upper bound:
\begin{eqnarray}
\| \hat \rho_{+} - \hat \rho_{-} \|_\tr &\leq&
 \min_{\pi \in S_{N_{\rm d}}} \frac{1}{N_{\rm d} } \sum_j^{N_{\rm d}} 
\left\| \hat \rho^{(j)}_{+} - \hat \rho^{(\pi(j))}_{-} \right\|_\tr \leq \nonumber \\
&\leq& \min_{\pi \in S_{N_{\rm d}}} \frac{1}{N_{\rm d} } \sum_j^{N_{\rm d}} 
\sqrt{1 - F_U \left( \hat \rho^{(j)}_{+} , \hat \rho^{(\pi(j))}_{-} \right)^2} \nonumber
\end{eqnarray}
The minimization over the set of permutation of $N_d$ elements can be done with the so-called ``Hungarian algorithm'' or ``Kuhn-Munkres algorithm''.~\cite{Munkres}
The computation of the Uhlmann fidelity $F_U$ between two mixed FGS has been explained in section~\ref{sec:FGstates}.

The fidelity of the optimal Gaussian operation requires the computation of the CM of $\hat \rho_{\pm}(t)$, which are not FGS. Because these states are convex combination of known FGS and because the CM is a linear function of the state, the CM of $\hat \rho_{\pm}(t)$ is:
\begin{equation}
 \Gamma_{\pm} (t) = \frac 1{N_{\rm d}} \sum_{j=1}^{N_{\rm d}} \Gamma_{\pm}^{(j)}(t)
\end{equation}
where $\Gamma_{\pm}^{(j)}(t)$ is the CM of $\hat \rho^{(j)}_{\pm}(t)$.

\section{Details on the Prior Knowledge of the Recovery Operation}\label{app:prior}

We complement the discussion of Sec.~\ref{sec:prior}.
We recall the definition of Uhlmann fidelity,~\cite{NC} $F_U(\hat \rho, \hat \sigma) \doteqdot \text{tr} \left[\sqrt{\sqrt{\hat \rho} \hat \sigma \sqrt{\hat \rho}}\right]$, which in case $\hat \rho$ is a pure state, $\ket{\Phi_{\hat \rho}}$, reduces to $F_U (\hat \rho, \hat \sigma) = \sqrt{ \bra{\Phi_{\hat \rho}} \hat \sigma \ket{ \Phi_{\hat \rho}}}$. Moreover, the following is true:~\cite{NC}
\begin{equation}
1 - F_U (\hat \rho, \hat \sigma) \leq
\frac 12 \| \hat \rho - \hat \sigma \|_{\text{tr}} \leq
\sqrt{1 - F_U (\hat \rho, \hat \sigma)^2} . 
\end{equation}

By definition $\| \mathcal R_t^{\op,(1)} (\hat \rho_1) - \hat \rho_q \|_{\rm op} < \varepsilon$, and $\varepsilon$ scales exponentially with the size $N$ and thus $ 1 \geq F_U (\mathcal R_t^{\op,(1)} (\hat \rho_1) , \hat \rho_q) \geq 1 - \varepsilon/2$.
Moreover, $F_U (\mathcal R_t^{\op,(1)} (\hat \rho_1) , \hat \rho_q)^2 \geq 1 - \varepsilon + \varepsilon^2 /4 \geq 1 - \varepsilon$.
Because $\text{tr}[\hat \rho_1 \hat \rho_2] \geq p$, where $p = |I_2|/|I_1|$, we can write $\hat \rho_1 = p \hat \rho_2 + (1-p) \hat \rho_3$, where $\hat\rho_3$ does not need to be better specified. Thus:
\begin{equation}
 F_U(\mathcal R_t^{\op,(1)} (\hat \rho_1) , \hat \rho_q)^2 \leq 
p F_U(\mathcal R_t^{\op,(1)} (\hat \rho_2) , \hat \rho_q)^2 + 1-p.
\end{equation}

From the derived inequalities, it follows that $F_U(\mathcal R_t^{\op,(1)} (\hat \rho_2) , \hat \rho_q)^2 \geq 1 - \varepsilon / p$. This leads to the final result:
\begin{equation}
\small{  \|  \mathcal R_t^{\op,(1)} (\hat \rho_2) - \hat \rho_q \|_{\rm tr}
\leq 
2 \sqrt{1 - F_U(\mathcal R_t^{\op,(1)} (\hat \rho_2) , \hat \rho_q)^2 }
\leq \frac{2\sqrt{\varepsilon}}{\sqrt p} .}
\end{equation}


\begin{references}
\bibitem{footnote3}
For any state, $\Gamma^T\Gamma\le \Id$ with equality only for pure Gaussian states. For any covariance matrix fulfilling this inequality, there exists a mixed Gaussian state corresponding to it.
\end{references}
\end{document}